**Developing a Comprehensive Model for Feasibility Analysis of Separated Bike Lanes and Electric Bike Lanes: A Case Study in Shanghai, China**

**Lu Ling**
PhD Candidate
Lyles School of Civil Engineering
Purdue University, West Lafayette, IN 47907
Email: ling58@purdue.edu

**Yuntao Guo, Ph.D.**
Assistant Professor
Department of Traffic Engineering and Key Laboratory of Road and Traffic Engineering
Ministry of Education, Tongji University
4800 Cao'an Road, Shanghai, 201804, China
Email: yuntaoguo@tongji.edu.cn

**Xiongfei Lai, Ph.D.**
Post-doctoral Research Fellow
Department of Civil and Environmental Engineering
National University of Singapore
1 Engineering Drive 2, #07-03, Singapore, 117576
Email: laixfcee@nus.edu.sg

**Tianpei Tang, Ph.D.**
Associate Professor
School of Transportation and Civil Engineering,
Nantong University, Nantong 226019, China
Email: tangtianpei@ntu.edu.cn

**Xinghua Li, Ph.D.**
Professor
Department of Traffic Engineering and Key Laboratory of Road and Traffic Engineering
Ministry of Education, Tongji University
4800 Cao'an Road, Shanghai, 201804, China
Email: xinghuali@tongji.edu.cn

Word Count: 6,200 words + 5 tables*250 = 7,450 words

*Submitted [July 31, 2021]*

*Lu Ling, Yuntao Guo, Xiongfei Lai, Tianpei Tang, Xinghua Li*

## ABSTRACT

Electric bikes (e-bikes), including lightweight e-bikes with pedals and e-bikes in scooter form, are gaining popularity around the world because of its convenience and affordability. At the same time, e-bike-related accidents are also on the rise and many policymakers and practitioners are debating the feasibility of building e-bike lanes in their communities. By collecting e-bikes and bikes data in Shanghai city, the study first recalibrates the capacity of the conventional bike lane based on the traffic movement characteristics of the mixed bikes flow. Then, the study evaluates the traffic safety performance of the mixed bike flow in the conventional bike lane by the observed passing events. Finally, this study proposes a comprehensive model for evaluating the feasibility of building an e-bike lane by integrating the Analytic Hierarchy Process and fuzzy mathematics by considering the three objectives: capacity, safety, and budget constraint. The proposed model, one of the first of its kind, can be used to (i) evaluate the existing road capacity and safety performance improvement of a mixed bike flow with e-bikes and human-powered bikes by analyzing the mixed bike flow arrival rate and passing maneuvers, and (ii) quantify the changes to the road capacity and safety performance if a new e-bike lane is constructed. Numerical experiments are performed to calibrate the proposed model and evaluate its performance using non-motorized vehicles' trajectories in Shanghai, China. The numerical experiment results suggest that the proposed model can be used by policymakers and practitioners to evaluate the feasibility of building e-bike lanes.

## INTRODUCTION

With the rapid expansion of urban scale in China, electric bikes (i.e., bikes with an electric or a gas-electric hybrid engine, hereafter referred to as "e-bikes") play a vital role in people's daily travel as well as last-mile delivery. E-bike is considered as a faster non-motorized travel mode compared to human-powered bike, a more affordable and environmentally friendly mode compared to private vehicle, and a more flexible mode for door-to-door travel compared to transit. These reasons make e-bike an attractive travel mode compared to other modes of transportation, particularly for those with low-to-medium income or living in neighborhoods with inconvenient public transportation systems (Guo et al., 2018; Lai et al., 2020a, 2020b; Ling et al., 2018; Ling and Li, 2017; Tang et al., 2021). The advantages of e-bike contribute to the rapid increase in e-bike ownership over the past two decades worldwide. For example, e-bike accounted for 75% of all bike mode share, and 19% of all trips in 2011 in Kunming city, China (Cherry et al., 2016) and the total number of e-bikes in China was 250 million in 2017 according to the Chinese cycling association (Zheng et al., 2017), which is a sharp contrast to the 58,000 in 1998. Besides, over 40 million e-bikes were sold in 2015 worldwide (Salmeron-Manzano and Manzano-Agugliaro, 2018) and the e-bike sale would rise to 130 million by 2025 and 800 million by 2100 (Jamerson and Benjamin, 2012). Two of the most commonly used e-bike types in China are (i) lightweight e-bikes with pedals (hereafter referred to as "pedal-assisted e-bikes") which have the option of operating by electricity or human-power and (ii) e-bikes in scooter form (hereafter referred to as "scooter e-bike") which are powered by a gas-electric hybrid engine as shown in **Figure 1**. The latter ones are often heavier but with a faster top operating speed and a longer operating range compared to the former ones. The safety standard for e-bikes and human-powered bike presents in **Table 1**(Electric travel, n.d.). In general, most of e-bikes in China have a top operating speed of around 30km/h, weight ranging from 30 to 80 kg, and consumes about 1.8 kWh/100 km (about one-tenth of an electric car)(Cherry et al., 2016; Ji et al., 2012). The operating speed and weight of human-powered bikes is around 15 km/h and 17 kg(Lin et al., 2008). In contrast, in accordance with European legislation, e-bikes have a legal limit of 25 km/h and a 250W motor output (Electric travel, n.d.).

Although much attention is paid to the vehicle crash frequency and severity(Ling et al., 2023; Ukkusuri et al., 2020), the rapid increase in e-bike ownership and usage brings much needed mobility to millions of travelers but also contributed to the sharp increase in e-bike-involved accidents in recent years. The annual e-bike-involved fatalities increased by more than tenfold and total number of e-bike-involved accidents increased by five times between 2004-2015 in China. Especially, there were over 11,000 e-bike-involved fatalities in China which represented around 90% of the non-motorized fatalities in 2015 (Jin et al., 2015) . In 2019, e -bikes accounted for 13.8% of all road traffic deaths and 17.4% of road traffic injuries in 2019 as reported by World Health Organization (World Health Organization, n.d.). From the severity point of view, the number of deaths increased by 71.52% in 2011 and 2016, and the numbers of e-bike injuries increased by 72.90% in five years (Ma, 2010). Although motor vehicles were the most often involved in the e-bike related crashes (56.5%), the crashes between e-bikes and motorcycle and bikes account for more than 17.6% of e-bike related crashes (T. Wang et al., 2018). The increasing e-bike-involved accidents can be partly contributed to the lack of classification standards and management strategies of e-bikes in China (Ji et al., 2012). That is, the size and the speed of e-bikes vary a lot in China, but they are classified and managed as human-powered bikes (e.g., limited regulations and enforcement on license, helmet, and insurance requirements). E-bikes were classified as the non-motorized vehicles and share the bike lane with human-powered bikes. This results in a mixed bike flow (human-powered bikes, pedal-assisted e-bikes, and scooter e-bikes) with different speed, size, and passing maneuver in bike lanes, which potentially increase the probability of having e-bike-involved accidents (T. Wang et al., 2018). In addition, recent studies (Jin et al., 2015; Lin et al., 2008) also found that e-bikers have a stronger tendency to violate traffic rules such as speeding and running a red light compared to other road users.

To address these safety challenges, city planners and researchers proposed to reconstruct traditional bike lanes and separate e-bikes from human-powered bikes by introducing dedicated e-bike lane(Cardoso et al., 2016). However, for such plan to come to fruition, it is important to analyze the level of service (e.g., safety and capacity benefits) of separating e-bikes and human-power bikes compared to the existing ones. However, most existing studies only focused on mixed flows of motor vehicles, pedestrians, and human-powered bikes. Leaving the mixed bike flow characteristics (speed, density, and passing maneuver) remains unexplored. In reality, e-bikes' movement characteristics, such as speed, acceleration, deceleration, and riders' behavior, are very different from traffic flows of other vehicles and pedestrians. Therefore, the study first analyzed the arriving headway of mixed bike flow (human-powered bikes, pedal-assisted e-bikes, and scooter e-bike), passing maneuver, and mixed bike lane capacity based on the observed data collected in 2016 in Shanghai, China. Then, a comprehensive evaluation method for dedicated e-bike lanes was developed by the Analytic Hierarchy Process (AHP) and fuzzy mathematics approaches. The approaches were applied to assess the level of service and evaluate the potential benefits of dedicated e-bike lanes in terms of improving lane capacity and rider safety within a budget constraint. Lastly, the numeric verification of the proposed evaluation method is made by Monte-Carlo simulation. Such method represents the first of its kind in evaluating the potential of dedicated e-bike lanes and can be used by planners and policymakers to explore the system-level benefits of constructing dedicated e-bike lanes.

The rest of the study is organized as follows. Section 2 briefly reviews related studies. Section 3 outlines the data used in our study. Section 4 introduces the methodology, which contains the theory basis of analytic hierarchy process and fuzzy dynamic analysis. Section 5 presents the results from the numerical experiment. on the analyses. Section 6 summarizes the major findings with concluding remarks.

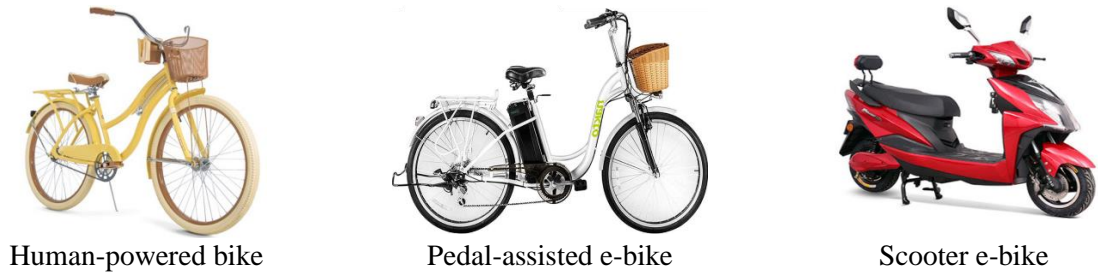

| Human-powered bike | Pedal-assisted e-bike | Scooter e-bike |

**Figure 1 Human-powered Bikes and E-bikes**

**Table 1. The Comparison of the Design Criterion**

|  | Human-powered bike | Pedal-assisted e-bike | Scooter e-bike |
|---|---|---|---|
| Design speed limits | ≤ 25 km/h | ≤ 25 km/h | ≤ 50 km/h |
| Motor Power | -- | ≤ 400w | 400w ~ 4kw |
| Weight | ≤ 55kg | ≤ 55kg | No restriction |
| Battery Voltage | -- | ≤ 48V | No restriction |

## LITERATURE REVIEW

Much attention has been devoted to the bike level of service (BLOS), which refers to the operating quality in the bike lane with only human-powered bike flow. This was first proposed in the BLOS at HCM 2000 (Haustein and Møller, 2016), which quantifies the safety and capacity of a bike lane. The criteria used in the BLOS is measured by a weighted sum of the number of passing and meeting events. There are multiple studies have explored the passing maneuvers for human-powered bikes. (Virkler and Balasubramanian, 1998) reported that the number of passing events is directly proportional to the standard deviation of flow speed and inversely proportional to mean speed based on the analyses of the exclusive bike lanes (only human-powered bikes used lane) and shared bike lanes (human-powered bikes shared with jogger and hikers). Later, (Khan and Raksuntorn, 2001) found that the human-powered bikes in the shared bike lane maintained a constant speed difference during the passing events and reported that the variation of speed difference during passing is one-third of variation for either the passing or the passed bicycle speed. Besides, they suggested that the average maximum lateral spacing during passing events and meeting events is 1.88 m. (Hummer et al., 2006) developed a new method by adding the adjustment of delayed passes to the BLOS equation. This method was later adopted in HCM 2010 (Manual, 2010). Besides, (Li et al., 2010)speculated that the passing events would increase if the bike speed increased. They also pointed out the free passing events (the horizontal distance of free space between overtaker and overtakee is more than the width of one lane) are more than the adjacent passing events (the horizontal spacing of free space is only enough for one bicycle to pass through) and delayed passing events (the horizontal spacing of free space is not enough for one bicycle to pass through) with the increasing of exclusive bike lane width. Under the assumption that human-powered bikes and e-bikes have similar movement characteristics (e.g., passing maneuver), the widely used description for an exclusive bike lane capacity still follows when there are only human-powered bikes in the lane (Chen et al., 2018).

Since the e-bike accounts for a small part of the daily commute in Europe and the United States, most of the studies in these regions focus only on the bike flow without factoring the impacts of e-bike on estimating the BLOS with mixed bike flow. Unlike these regions, e-bike presents one of the key modes of transportation in many Asian countries, and mixed flow of human-powered bikes and e-bikes are causing rising safety issues. It is one of the pressing issues faced by many policymakers and planners. One of the main reasons for accidents in the mixed bike flow is the speed differences during passing maneuvers among human-powered bikes, pedal-assisted e-bikes, and scooter e-bike. As there are speed differences among these three types of two-wheelers, the number of passing events in mixed bike flow is expected to be much larger compared to that of bike flow with only human-powered bikes. For example, (Lin et al., 2008) found that the mean operating speed of e-bike is 47.6% higher than that of human-powered bikes and that most e-bike riders ride faster than the bike lane speed limit. In addition, (Minh et al., 2005) suggested that the speed difference of the passing motorcycle and passed motorcycle is 6.7km/h, and average speed of the vehicle in pair is positively related to the lateral clearance with a correlation coefficient of 0.46 for exclusive motorcycle lane based on the observed data in Vietnam. (Lin et al., 2014) suggested that the lateral spacing would decrease during passing events and the relationship of the lateral spacing during passing is 'moped e-bike passing human-powered bikes' having the largest spacing, followed by 'moped e-bikes passing moped e-bikes', and 'human-powered bikes passing human-powered bikes' having the smallest spacing based on the observations in the shared bike lane (human-powered bikes shared with the moped e-bikes in Shanghai, China. (Zhao et al., 2013) and (Xu et al., 2018) found the number of passing events would first increase and then decrease as the ratio of e-bike increases from 0 to 100 in the mixed bike flow. When the speed differences between bikes and e-bikes are small, it can also lead to overtake disturbances (the lateral acceleration volatility during pass events) which can be dangerous for the riders. (Chen et al., 2018) suggested that both widening the bike lane and posting a speed limit could mitigate the passing events for the mixed bike flow based on the observations in Shanghai. Besides, the number of illegal lane-changing behavior (i.e., non-motorized vehicles occupy the adjacent vehicle lane while traveling) increases with the rise of the proportion of e-bikes (Li et al., 2020) . In addition, (Liu et al., 2020) suggested that the passing events between human-powered bikes and e-bikes reach the maximum when the share of e-bikes is 70% on the roads shared by motorized and non-motorized vehicles.

The most common type of methods used to evaluate the BLOS for mixed bike flow is the microsimulation approaches which require the traffic flow characteristics as inputs. Among these methods, cellular automata (CA) model is one of the most commonly used one in analyzing the mixed bike flow (Jia et al., 2007; Jiang et al., 2004). The CA model separates the road segment into cells with fixed size to accommodate the human-powered bikes or e-bikes and update the individuals based on the speed and location evolution rules. Later, (Shan et al., 2015) proposed the modified CA model to study the mixed bike flow involving e-bikes. The modified CA also indicates that a smaller proportion of e-bikes may cause stability concerns due to large speed differences between human-powered bikes and e-bikes (Zhou et al., 2015). However, the CA model two key limitations. On one hand, it cannot capture the condition when the actual flow exceed the capacity. On the other hand, it cannot capture the human behaviors based on traditional grid and cell size. Social force model (SFM) has emerged as alternative microsimulation model by cooperating the human behavior to simulate the movement characteristic for the mixed flows among cars (Qu et al., 2017), pedestrians (Ning-bo et al., 2017), and bikes (Huang et al., 2016)  in recent years. SFM models bike flow in a continuous way and updates the magnitude and direction of the speed model according to the received forces at any time (Ning-bo et al., 2017). Thus, the simulation results are more fit with practical traffic flow. However, the SFM is complicated with diverse parameters in formulating the social force and speed variations. Furthermore, this model has yet to be applied to study mixed bike flows with human-powered bikes and e-bikes.

Despite that the aforementioned studies provide valuable insights for understanding the movement characteristic of the mixed bike flow and developing various types of microsimulation models to evaluate the BOLS of mixed bike flow, there are two key limitations that have yet to be addressed. On one hand, most studies only focused on the speed difference between bikes and e-bikes but have yet to incorporate other behavioral differences between them in the model. On the other, most existing models did not separate pedal-assisted e-bike and scooter e-bike and assumed that both are same type of e-bikes. In reality, they are difference in terms of speed, quality, acceleration/deceleration, and driver behavior. Thus, the bias would be introduced if separating them as two classes (human-powered bikes and e-bikes). Without addressing these limitations, it can be challenging to provide a comprehensive understanding in terms of the impacts of exclusive e-bike lanes on both rider safety (passing maneuvers of the mixed bike flow) and bike lane efficiency (the capacity of the bike lane).

## DATA COLLECTION AND ANALYSIS

To understand the movement characteristic of the mixed bike flow in the bike lane, it is important to first study the trajectory of the mixed bike flow on bike lanes. One of the most important challenges is to measure the average headway of mixed bike flow. The movement of the non-motorized vehicles are less restrictive compared to motorized vehicles, and no distinct lane for each bike/e-bike as shown in **Figure 2**. To get a more accurate measurements, we used video recordings of the non-motorized vehicle movements from two intersections on four roads (i.e., West Nanjing Rd, Changde Rd, West Tianmu Rd, Datong Rd) with exclusive bike lane in Shanghai, China. The video recording for each intersection is thirty minutes recorded at peak hours and the headway measurement is during e-bikes queue discharging. The selected roads have high e-bike volumes as shown in **Figure 3**. We defined nine types of non-motorized vehicle headway in the study, including scooter e-bike followed by scooter e-bike (M-M), scooter e-bike followed by pedal-assisted e-bike (M-E), scooter e-bike followd by human-powered bike (M-B), pedal-assisted e-bike followed by scooter e-bike (E-M), pedal-assisted e-bike followed by pedal-assisted e-bike (E-E), pedal-assisted e-bike followed by human-powered bike (E-B), human-powered bike followed by scooter e-bike (B-M), human-powered bike followed by pedal-assisted e-bike (B-E), human-powered bike followed by human-powered bike (B-B). Finally, we collected 572 following behavior of mixed bike for nine types of headway.

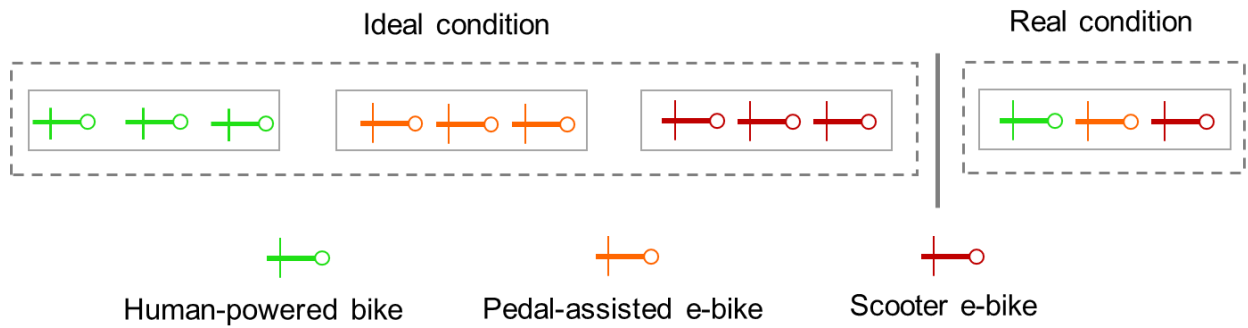

**Figure 2 Human-powered Bikes and E-bikes Flow**

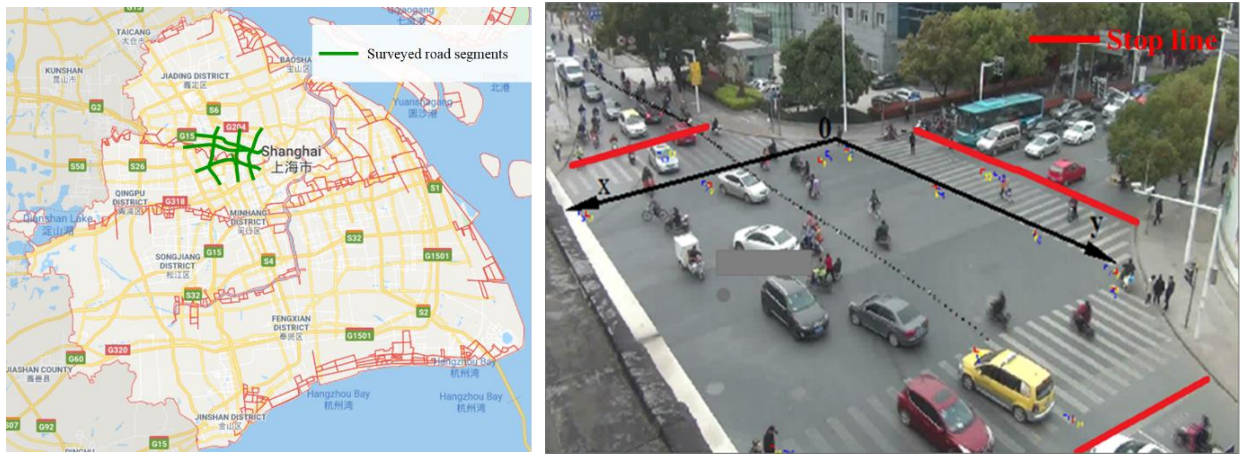

**Figure 3 Study Area and Data Acquisition**

**Arrival Rate of Mixed Bike Flow**

Based on the queneing theory , the time series of two consecutive unit time follow the negative exponential distribution of the same parameter if the arrival rate of vehicles per unit time follow Poisson distribution. Two assumptions were made related to the arrival rate of the non-motorized vehicle flow:

(1) the mixed bike flow arrival process follows the Poisson distribution, as shown in the **Equation 1**:

$$P(X = x) = \frac{(\lambda x)^x e^{-\lambda T}}{x!} \tag{1}$$

where $P(X = x)$ is the probability of event $X$ that repeats $x$ times during the observation period $T$ ; event $X$ is each type of headway (M-M, M-E, M-B, E-M, E-E, E-B, B-M, B-E, B-B); $\lambda$ is the average number of events $X$ per interval; $T$ is the observed period.

(2) the grouped headway of mixed bike flow follows the normal distribution.

We assume the arrival events of the non-motorized vehicles are identical and independent, and we randomly selected the arrival process of the scooter e-bikes, pedal-assisted e-bike, and human powered bikes). We then examined the hypothesis that 'the grouped headway of the non-motorized vehicle follows normal distribution' by z-test . The formula of normal distribution can be written in **Equation 2**:

$$f(x) = \frac{1}{\sigma\sqrt{2\pi}} e^{-\frac{1}{2}\left(\frac{(x-\mu)^2}{2\sigma^2}\right)} \tag{2}$$

where f(x) is a normal distribution function; $\mu$ is the average headway for each type of non-motorized vehciles; $\sigma$ is the standard deviation. The data collected is used for hypothesis testing and the results are shown in **Table 1**.

**TABLE 1 Headway Hypothesis Testing (Evening peak hours)**

| Group | Average(s) | Standard Deviation | P-value | Conclusion |
|-------|-----------|--------------------|---------|------------|
| M-M | 1.174 | 0.41 | 0.707 | Fail to reject |
| M-E | 1.322 | 0.64 | 0.321 | Fail to reject |
| M-B | 1.570 | 0.53 | 0.949 | Fail to reject |
| E-E | 1.606 | 0.35 | 0.294 | Fail to reject |
| E-M | 1.346 | 0.45 | 0.808 | Fail to reject |
| E-B | 1.643 | 0.57 | 0.976 | Fail to reject |

| | | | | | |
|---|---|---|---|---|---|
| B-B | 1.740 | 0.38 | 0.667 | Fail to reject |
| B-M | 1.349 | 0.34 | 0.270 | Fail to reject |
| B-E | 1.499 | 0.26 | 0.989 | Fail to reject |

*Note: "M" means scooter e-bike, "E" means pedal-assisted e-bike, "B" means human powered bike；

From the hypothesis testing, the headway distribution follow a normal distribution. Besides, the M-M has the smallest headway and B-B has the largest headway, and M-E, E-M, B-M, and B-E have relatively shorter headway than the headways of E-E, M-B, E-B. That is because the scooter e-bikes have the highest operating velocity and accelerated velocity among three types of non-motorized vehicle and human-powered bikes usually are slower than others, which is consitent with the previous findings (Zhao et al., 2013). However, E-E has larger headway than M-B. Because human-powered bike riders feel less pressure when they follow with the scooter-ebikes compared due to the condifence that the scooter-ebikes have a much higher operating velocity than themselves. While the uncertainty would increase if the operating velocity is the same for two pedal-assisted e-bikes in the following behavior. And that might lead to a larger longitudinal and lateral spacing for E-E than M-B.

**Traffic Capacity**
According to the Large Number Theorem , the expected value is approximate to the observed mean from a large number of events if the events is identical and independent distributed and follows a normal distribution. Therefore, we adapted the observed mean headway of each grouped as the expected headway of each group, and estimated the capacity of bike lane for mixed bike flow. The estimated average headway in four observed roads can be found in **Table 2**. Therefore, the capacity of bike lane with mixed bike flow can be seen in **Equation 3**:

$$\bar{t} = t_{B-B}x^2 + t_{E-E}y^2 + t_{M-M}y^2 + 2(t_{B-E} + t_{E-B})xy + 2(t_{B-M} + t_{M-B})xz + 2(t_{E-M} + t_{M-E})yz \ (3)$$

where $\bar{t}$ is the estimated headway for a bike lane with mixed bike flow;

$x$ is the percentage of the human-powered bike flow ratio;

$y$ is the percentage of the pedal-assisted e-bike flow ratio;

$z$ is the percentage of the scooter e-bike flow ratio;

$t_{B-B}, t_{E-E}, t_{M-M}, t_{B-E}, t_{E-B}, t_{B-M}, t_{M-B}, t_{E-M}, t_{M-E}$ represent nine combinations of respective headway.

**TABLE 2 The Headway of Types of Non-motorized Vehicles in Each Intersection**

| Location | M-M(s) | M-E(s) | M-B(s) | E-E(s) | E-M(s) | E-B(s) | B-B(s) | B-M(s) | B-E(s) |
|---|---|---|---|---|---|---|---|---|---|
| Changde Rd. | 1.8 | 1.81 | 1.92 | 1.76 | 1.57 | 2.08 | 2.62 | 2.21 | 1.85 |
| West Tianmu Rd. (Section A) | 1.4 | 1.3 | 1.48 | 1.3 | 1.69 | 1.3 | 1.82 | 1.97 | 1.23 |
| West Tianmu Rd. (Section A) | 1.26 | 1.4 | 1.58 | 1.61 | 1.52 | 1.65 | 1.94 | 1.53 | 1.64 |
| West Tianmu Rd. (Section B) | 1.13 | 1.3 | 1.49 | 1.27 | 1.5 | 1.15 | 1.81 | 2.02 | 1.79 |
| West Tianmu Rd. (Section B) | 1.21 | 1.57 | 1.49 | 1.33 | 1.44 | 1.55 | 1.49 | 1.58 | 1.28 |
| Average | 1.36 | 1.48 | 1.59 | 1.45 | 1.54 | 1.55 | 1.94 | 1.86 | 1.56 |

Using the data from **Table 2**, the non-motorized vehicle traffic capacity is in **Equation 4**:

$$N_{red} = \frac{3600}{1.94x^2 + 1.45y^2 + 1.36x^2 + 6.22xy + 6.9xz + 6.04yz} \tag{4}$$

The headway distribution of each types follows a normal distribution, and the standard deviation of the headway is smaller than mean, which means the data is not over-dispersion. We then apply the estimated headway for a bike lane with mixed bike flow into the standard bike lane capacity function. The corrected capacity of a bike lane with mixed bike flow can be seen in **Equation 5**, and the calculated actual capacity for four roads is presented in **Table 3**.

$$N_{ideal} = \frac{3600}{\bar{t}} \; veh/h/m \tag{5}$$

**TABLE 3 Corrected Capacity and Traffic Volume**

| Location | Capacity (Veh/h·m) | $P_1$ | $P_2$ | $P_3$ |
|---|---|---|---|---|
| Changde Rd. | 2404 | 49 | 24 | 109 |
| West Tianmu Rd. (Section A) | 2654 | 21 | 33 | 191 |
| West Tianmu Rd. (Section B) | 2566 | 22 | 37 | 121 |
| West Tianmu Rd. (Section A) | 2606 | 34 | 46 | 223 |
| West Tianmu Rd. (Section B) | 2586 | 21 | 41 | 135 |

Note: $P_1$ is the human-powered bike flow, $P_2$ is the pedal-assisted e-bike flow, $P_3$ is the scooter e-bike flow.

In light of the high correlation between the corrected capacity of bike lane awith the mixed flow of human-powered bike, pedal-assisted e-bike and scooter e-bike, the linear regression is in the **Equation 6**:

$$N_{real} = 2580 - 5.83P_1 - 0.558P_2 + 1.123P_3 \tag{6}$$

**Equation 6** presented the relationship between the capacity and actual traffic flow under the condition of a standard bike lane with the proportion of scooter e-bikes, pedal-assisted e-bikes, and human-powered bikes to be: 3:1:1.

## Traffic Safety Characteristics

Mobility and activity are key aspects to understand the traffic flow characteristics(Ling et al., 2022, 2021). In particular the safety measure is critical in evaluating the BLOS. Studies (Dong et al., 2015; Mehta et al., 2015) use the number of passing events reflect the safety level of the BLOC due to the fact that the more frequent the passing events occur, the higher collision probabliity. Therefore, "passing frequency" can be used to measure the possible risk-level of the mixed bike flow in the bike lane. In this study, we used "passing frequency" to measure the safety performance of the bike lane. The passsing frequency in surveyed segments was summarized in **Table 4**:

**TABLE 4 Survey Results of Passing Frequency (17:00pm -18:00pm)**

| Location | M-M | M-E | M-B | E-M | E-E | E-B | B-M | B-E | B-B | Number of lanes |
|---|---|---|---|---|---|---|---|---|---|---|
| West Tianmu Rd. (West-East) | 56 | 37 | 140 | 7 | 1 | 42 | 0 | 0 | 6 | 2 |
| West Tianmu Rd. | 50 | 10 | 48 | 0 | 0 | 1 | 0 | 0 | 6 | 2 |

| | | | | | | | | | |
|---|---|---|---|---|---|---|---|---|---|
| (East -West) | | | | | | | | | |
| Datong Rd. Tunnel (North-South) | 44 | 26 | 39 | 6 | 2 | 12 | 0 | 0 | 2 | 3 |
| Datong Rd. Tunnel (North-South) | 59 | 16 | 34 | 5 | 1 | 6 | 1 | 2 | 2 | 3 |
| West Tianmu Rd*. | 0.0121 | 0.0222 | 0.0229 | 0.0001 | 0.0043 | 0.0203 | 0.0000 | 0.0000 | 0.0031 | 2 |
| Datong Rd. Tunnel* | 0.0088 | 0.0038 | 0.0129 | 0.0010 | 0.0049 | 0.0105 | 0.0002 | 0.0012 | 0.0045 | 3 |

*: refers to the normalized passing frequency, which is calculated as the number of passings / (flow*flow)

As shown above, the number of scooter e-bike passing events is significantly higher than the passing events of others. Besides, we obtian several insights based on the average normalized passing frequency for two roads. The normalized passing frequency for M-B, E-B, and M-E are higher than other types. Among these three, the M-B has the highest normalized passing frequency. The scooter e-bikes naturally have the highest operating velocity than others would account for this observation. This fact also indicates that the scooter e-bike would greatly increase the passing frequency in the bike lane. Besides, we find that the normalized passing frequency for M-M is higher than E-E and B-B. One possible reason might be because the scooter e-bike users are more likely to have aggressive driving behavir, which is consistent with (C. Wang et al., 2018) . Those findings show that the passing frequency is influened by the types of non-motorized vehicle in the mixed-bike flow and the proportion for these non-motorized vehicles. Similar to the findings of (Xu et al., 2018), if the mixed bike flow is less than the bike lane capacity, as the percentage of the scooter e-bike riders in the mixed bike flow increase, the number of the passing events increases. Otherwise, the number of passing events remains the same despite the change in the percentage of the scooter e-bike riders. Thus, there is a causality relationship between the proportional of mixed bike flow and passing frequency. Based on these observations, we regressed the passing frequency with the traffic flow of scooter e-bikes, pedal-assisted e-bikes and human-powered bikes by OLS model (see **Equation 7**). Each component is significant at 0.05 level and the adjusted $R^2$ is 0.82. The function between the passing frequency and the proportion of mixed bike flow can be found below:

$$N_{passing} = 18.264 + 0.197P_1 - 2.268P_2 + 0.683P_3 \qquad (7)$$

**Equation 7** subjects to when the mixed bike flow is under the capacity of bike lane, and the traffic flow in the bike lane with the proportion of scooter e-bikes, pedal-assisted e-bikes, and human-powered bikes to be: 3:1:1. Since the speed difference between the scooter e-bikes and human-powered bikes is significant, the more scooter e-bikes and human-powered bikes possess, the higher the overtaking frequency occurs. As a result, the number of passing events is positively correlated with bike flow and scooter e-bike flow. Meanwhile, the pedal-assisted e-bikes can coordinate with the other two types of vehicles, so it is inversely proportional to number of passing events.

## METHODOLOGY

To assess the BLOS with mixed bike flow, we construct two scenarios. One scenario is to assess the BLOS under the current bike lane from the aspects of efficiency and safety, which is the recalibrated capacity and passing frequency in the study. The other one is based on the 'what-if' scenario in which the proposed e-bike lane with the same measures was assessed. For cost-benefit evaluation of each scenario, we evaluate the BLOS for the current bike lane with the recalibrated capacity function and passing frequency function for the first case. Besides, we consider the traffic efficiency, safety, and economy to assess the level of service of the proposed e-bike lane (ELOS). We establish a dynamic countermeasure evaluation model to compare the current case and the alternative cases based on the Analytical Hierarchy

Process (AHP) framework. The traffic efficiency $y_1$, safety $y_2$, and inherent economy $y_3$ were selected as evaluation indexes in the AHP. Traffic efficiency refers to the lane capacity $y_{1,1}$, which is a quantitative index. The passing frequency is used to describe the safety performance of the lane and is represented by $y_{2,1}$ of the mixed bike flow in the lane. Economy factor refers to the cost of road construction $y_{3,1}$ and maintenance $y_{3,2}$, which are used as the qualitative index.

**Comprehensive Evaluation System Established by AHP**

AHP(Vinay et al., 2013) is a multi-criteria decision-making method, which combines both the qualitative and quantitative method. It is a hierarchy model based on the levels in the subjective judgment, decision makers, and the weighted processing. It is one of the most effective methods to solve the multi criteria decision-making problem and select alternative.

According to the principle of AHP, the framework of evaluated index system can be constructed in **Figure 4**:

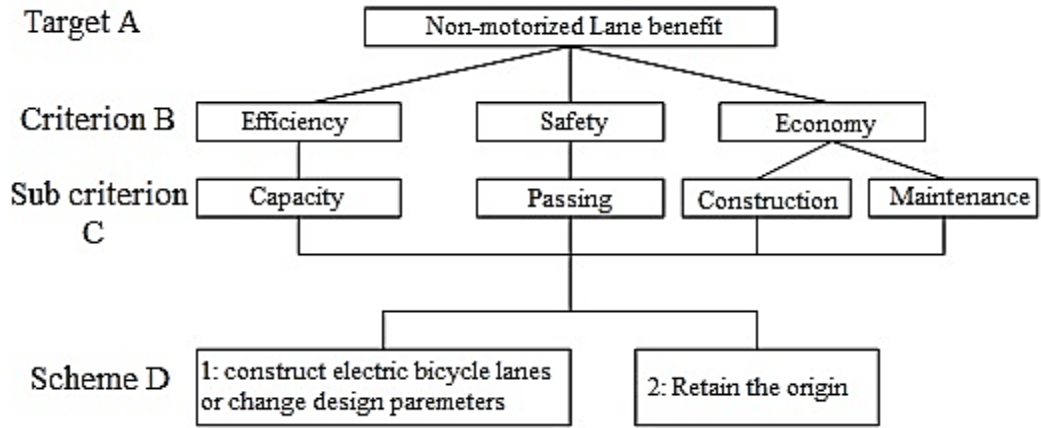

**Figure 4 Analytic Hierarchy Process Index System Construction**

To improve the safety of non-motorized vehicle and road efficiency. this paper establishes a bi-level evaluation system to analyze whether e-bike lanes should be added, or traditional bike lanes should be reconstructed. Theoretical steps are described as follows:

*Step 1*: Constructing comparison matrix. We first analyze the effects of the $n$ factor $y$ on target $A$, so that

$$W_{ij} = \frac{y_i}{y_j}, (i, j \in n) \tag{8}$$

**Equation**. 8 represents the ratio of $y_i$ and $y_j$ to $A$ and

$$Z = \begin{bmatrix} y_{1,1} & y_{1,2} \\ y_{2,1} & y_{2,2} \\ y_{3,1} & y_{3,2} \end{bmatrix} = \begin{bmatrix} \dfrac{y_1}{y_1} & \dfrac{y_1}{y_2} \\ \dfrac{y_2}{y_1} & \dfrac{y_2}{y_2} \\ \dfrac{y_3}{y_1} & \dfrac{y_3}{y_2} \end{bmatrix} \tag{9}$$

$Z$ is a pairwise comparison matrix on a 1-9 gradient scale.
*Step 2*: Consistency check of judgment matrix. According to the steps of analytic hierarchy process,

$$V = (y_1, y_2, \cdots, y_i)^T, i \in n \tag{10}$$

where $V$ is a unit characteristic vector. $S$ is the judgment matrix, and the value of the $W$ is normalized as the ranking weight of factors, and the maximum characteristic matrix of the judgment matrix is:

$$\lambda_{MAX} = \sum_{i=1}^{n} \frac{(SV)_i}{nV_i}, n = 3 \tag{11}$$

Using $(\lambda_{MAX} - n)$ as the consistency index, the expression is $CI = \frac{\lambda_{MAX} - n}{n-1}$, and $n$ is the order of the judgment matrix. $RI$ represents random consistency index which have the same order as $CI$. $CR = \frac{CI}{RI}$ is consistency ratio, and if CR<0.1, it is generally considered that the degree of inconsistency of $S$ is within the allowed range, and its eigenvalues can be used as weight vectors. Otherwise, $S$ needs to be adjusted until the $Step$ conformance test meets the requirements.

*Step 3*: Determining the weight vector. The upper level $A$ contains $n = 3$ factors: $A_1, A_2, A_3$, whose hierarchical total sorting weights are $y_1, y_2, y_3$. The lower level $B$ contains $m = 4$ factors: $y_{1,1}, y_{2,1}, y_{3,1}, y_{3,2}$, and their single order weights for $A_n (n=1,2,3)$ are denoted as $W_{1n}, W_{2n}, \ldots, W_{mn}$. At this point, the total weight of $B$ layer is:

$$w_m = \sum_{j=1}^{n} y_j x_{mj}, m = 4, n = 3 \tag{12}$$

## Fuzzy Dynamic Analysis Model

In large-scale problems with many criteria and subjective evaluation, fuzzy mathematics is a good way to avoid subjectivity of AHP. By introducing a dynamic membership, the membership size will be classified and used for the quantitative analysis for fuzzy real and evaluate the estimator. This is the basic idea of solving the problem with fuzzy mathematics theory (Zadeh et al., 1996).

The comprehensive judgment of fuzzy mathematics mainly includes 4 aspects: factor set $y$, scheme set $D$, subordinate matrix $R$, and weight vector $W$ (which has been given by analytic hierarchy model). From the fuzzy mathematics theory, the factor set $y$ has the reaction value on the selected $D = \{D_1, D_2\}$, $D_1$ means constructing e-bike lane or reconstructing bike lane; $D_2$ means remaining the tradition bike lane. The weight value $W$ is given by AHP, which should meet the requirement that $0 < W_{mn} < 1$, and $\sum M = 1$. The membership degree of quantitative indicators can be determined by the membership function method, in which the quantitative indicators can be divided into consumability and profitability indicators. For profitability indicator, the larger the index, the better. For consumptive indicators, the smaller the index, the better. Among them, the formula of the profit index can be expressed as $\mu_{i,j} = \frac{y_{i,j}}{\max(y_{i,j})} (i = 1,2,3; j = 1,2)$, and the consumptive index can be expressed as $\mu_{i,j} = \frac{\min y_{i,j}}{y_{i,j}} (i = 1,2,3; j = 1,2)$. For qualitative indicators, two-factors comparison method can be adopted. The factors are judged by the degree of contribution to the target: If $y_{1,2}$ is more important than $y_{2,1}$, it makes $e_{1,2} = 1, e_{2,1} = 0$ if equal importance, $e_{1,2} = e_{2,1} = 0.5$. $e_{i,j}$ is the fuzzy scale of the importance of $i$ to the target $j$, and $e_{j,i}$ is the importance fuzzy scale of the target $j$ to the target $i$. Finally, the relative membership degree of the qualitative index can be obtained according to the query scale of the mood operator (Lee, 2004).

## Dynamic Comprehensive Evaluation Model

Since the non-motorized lane's evaluation is influenced by three kinds of non-motorized participants, the evaluation model depends on the membership matrix and weight matrix based on fuzzy mathematics. It means that the comprehensive superiority makes judgment according to the principle of maximum membership degree. It can be calculated as follows,

$$D = MAX(D_j); j = 1, 2 \tag{13}$$

where $D$ means Scheme selection; $D_1$ means reconstruct the bike lane or add e-bike lane; $D_2$ means remaining the traditional bike lane.

According to the section of the evaluation index system, human-powered bikes, pedal-assisted e-bike and scooter e-bike have a very close contact with e-bike lane. Therefore, this paper establishes a dynamic evaluation model based on bike flow, pedal-assisted e-bike, and scooter e-bike flow. With the change of non-motorized vehicle flow, the weights of influencing factors of evaluation indexes will change, which will affect evaluation results and decisions of scheme layer.

The comprehensive evaluation function of the scheme layer can be written as:

$$D_j = \sum_{y_i=1}^{3}[w_{y_i} g f_{y_i}(\mu_{j,y_i})]$$
$$= \sum_{y_i=1}^{3}\{w_{y_i} g \sum_{y_j=1}^{2}[w_{y_{i,j}} f^{'}_{y_{i,j}}(\mu^{'}_{j,y_{i,j}}(P_1, P_2, P_3))]\}$$

(14)

where $w_{y_i}$ is the weight matrix of the first level evaluation index; $f_{y_i}(\mu_{j,y_i})$ represents the evaluation value of the index $y_i$, which is determined by the weights of the two indicators and the evaluation value; $\mu_{j,y_i}$ is the membership function index, which is influenced by bike flow $P_1$, pedal-assisted e-bike flow $P_2$ and scooter e-bike flow $P_3$.

Similarly, $w_{y_{i,j}}$ represents the weight matrix of the second level evaluation index $y_{i,j}$ relative to the first level index $y_i$. $f'_{y_{i,j}}(\mu'_{j,\ y_{i,j}})$ indicates the evaluation value of the second level index $y_{i,j}$, and $\mu'_{j,\ y_{i,j}}$ represents the two-grade membership function. In this paper, the influence factor of the mood factor can be calculated as the dimensionless principle by fuzzy mathematics principle. Finally, optimal scheme can be calculated according to the principle of maximum membership degree.

## NUMERICAL EXPERIMENTS

### Calibration of Membership Degree

*Traffic efficiency index*

Based on the video data acquisition in Shanghai, we calibrated each index in the model. According to **Equation 6**, the real capacity of a bike lane change with the different volume combination of human-powered bikes, pedal-assisted e-bikes, and scooter e-bikes. Since traffic capacity is a profit index, the larger the traffic capacity is, the greater the road benefit is. The membership function can be expressed as:

$$\mu_{i,j} = \frac{y_{i,j}}{\max y_{i,j}}(i=1,2,3; j=1,2)$$

(15)

The value of membership function is between 0~1, and according to **Equation 6** when it is an e-bike lane, the maximum traffic capacity is 2858Veh/h·meter. Therefore, according to the principle of fuzzy mathematics, the membership function can be formulated as:

$$\mu_{i,1} = \frac{2580 - 5.83P_1 - 0.558P_2 + 1.23P_3}{2858}$$

(16)

*Safety index*

Safety index was determined by the conflict possibility of non-motorized vehicles, in which the passing frequency is well-described. Passing frequency is proportional to the three-party flow, and also is the profit index. Therefore, with the same situation as capacity index, it can be descripted as,

$$\mu_{i,j} = \frac{y_{i,j}}{\max y_{i,j}}(i=1,2,3; j=1,2)$$

(17)

The maximum passing number of one bike lane is 150 veh/h/lane. On that basis, the membership function is defined as:

$$\mu_{i,2} = \frac{18.284 + 0.197P_1 - 2.268P_2 + 0.682P_3}{150} \tag{18}$$

*Economy indicators*

There is no construction and maintenance cost for the new project in current case. Thus, the economy indicator in the criterion B of AHP is 0. For the alternative case, a new e-bike lane is constructed. Therefore, the economy indicator with construction and maintenance cost should be considered in the 'what-if' scenario. Since there is no actual financial estimator in the scheme, the qualitative membership function is adapted in the study, and the two-factor comparison method is adopted to calibrate it. That is, we can use the tone operator "incomparable" class to compare the two schemes, so the relative membership degree is:

$$\mu_{1,3,1} = 1, \quad \mu_{2,3,1} = 0 \tag{19}$$

Maintenance cost is basically consistent with construction cost principle, and maintenance cost is only relative to the traffic volume of non-motorized vehicle. Assuming that the choice of the schemes will not affect individual's travel behavior. So, non-motorized vehicle flow will not change and maintenance costs of the two schemes are almost the same:

$$\mu_{1,3,2} = 0.5, \quad \mu_{2,3,2} = 0.5 \tag{20}$$

**Calibration of Index Weight**

*Weights of the first factors layer*

The three evaluation criteria of the first level are different in the two schemes. In the first scheme where it supports reconstruction of the bike lane, more attention was paid to the safety of bike and the improvement of road efficiency, while the economic considerations were relatively tolerable. In the second scheme where it suggests remaining the original bike lane, the economic factors are more important, and the tolerance threshold is relatively higher than the safety index and the efficiency index.

According to the analytic hierarchy process, the relative scales for the first scheme and second scheme are shown in the **Table 5**:

**TABLE 5 Relative Scaling for Two Schemes**

| Scheme | Relative scale | Efficiency | Safety | Economy |
|--------|---------------|------------|--------|---------|
| | Efficiency | 1 | 1/2 | 2 |
| 1 | Safety | 2 | 1 | 4 |
| | Economy | 1/2 | 1/4 | 1 |
| | Efficiency | 1 | 1/2 | 1/4 |
| 2 | Safety | | 1 | 1/2 |
| | Economy | 4 | 2 | 1 |

Where, $n = 3$, $RI = 0.58$, and eigenvalue is: $\lambda = 3$, so $CI = \frac{\lambda - n}{n-1}$ and $CR = \frac{CI}{RI} = 0$, it conforms to the consistency test principle for two schemes. The parameter calibrated in the first scheme: $j = 1, w_{j,1} = \frac{2}{7}, w_{j,2} = \frac{4}{7}, w_{j,3} = \frac{1}{7}$ and $j = 2, w_{j,1} = \frac{1}{7}, w_{j,2} = \frac{2}{7}, w_{j,3} = \frac{4}{7}$ in the second scheme.

*Weight of the second factors layer*

As for the second level indexes, the traffic efficiency is measured only by traffic capacity, so its weight is 1. The safety performance is captured by passing frequency, which is 1 relative to the previous layer. The level of financial budget has two factors: construction cost and maintenance cost. We consider the construction cost as two times relative to maintenance cost. The consistency check is $CR$=0 based on AHP, which conform to the requirements of conformance testing. Therefore, the weights of two economic factors are: $w_{j,3,1} = \frac{2}{3}, w_{j,3,2} = \frac{1}{3}, (j = 1,2)$.

**Numerical Verification of the Dynamic Evaluation Model**

The above parameters are brought into function 10, and the fuzzy evaluation index of grade one and two can be obtained. The final expression of fuzzy function is:

$$D_j = [(0.4441 + 1.94 \times 10^{-4} P_1 - 0.5665 P_2 + 0.00268 P_3), (0.63871 + 9.54 \times 10^{-5} P_1, -0.56228 P_2 + 0.00134 P_3)], (j = 1,2) \quad (21)$$

where $P_1$, $P_2$, $P_3$ denote the flow of human-powered bikes, pedal-assisted e-bikes, scooter e-bikes.

According to the principle of maximum degree of membership, we evaluate which scheme should be selected when the flow of bike, pedal-assisted e-bike and scooter e-bike are defined. According to the principle of maximum membership degree, we adapt Mont-Carlo algorithm to simulate the flows combination, given a series of random volumes (which under the model definition). The evaluation threshold of the transition section is set to 0.05, and the result is shown in the figure:

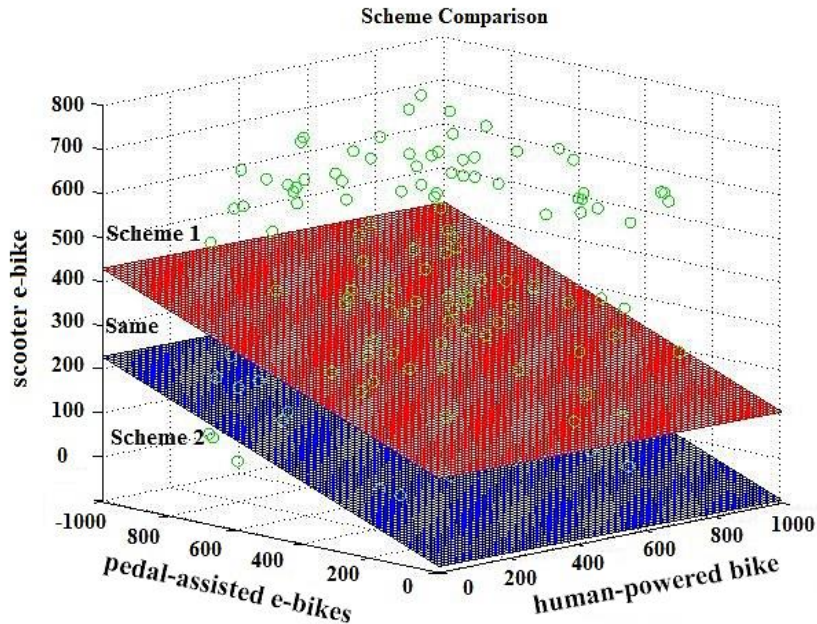

**Figure 5 The Numerical Verification of Dynamic Valuation Model**

**Figure 5** shows that scheme 2 is the optimal selection when the flow rate of pedal-assisted e-bike or scooter e-bike is small. And scheme 1 is the optimal selection when the flow rate of pedal-assisted e-bike or scooter e-bike is large. At the same time, if pedal-assisted e-bike or scooter e-bike flow is in the middle, the evaluation results are similar, and the selection model is moderate. From the numerical verification, two schemes have a relative moderated gap, which requires more evaluation indexes to help improve the selected model. Besides, bike flow seems has little influence on the scheme selection, which just as the analysis mentioned before.

**CONCLUSIONS**

Although multiple studies confirmed the impact factors for bike safety problems, few studies explored the movement characteristics of mixed bike flow. This paper conducts a detailed movement characteristic for mixed bike flow based on field survey data. Then, we discuss the applicability of e-bike lane under different non-motorized traffic flow combination from 3 aspects: traffic efficiency, safety and economy. Finally, on the basis of schemes about whether to reconstruct bike lane or to remain the original bike lane, a dynamic evaluation function model based on AHP, and fuzzy mathematics is built to select the optimal scheme on certain proportion of non-motorized vehicles. This model is calibrated based on the video data of bike lane in Shanghai and the optimal scheme of the model is selected by numerical verification. This study not only puts forward a basic understanding for traffic characteristics of non-motorized vehicles, but also gives suggestions to industrial managers and evokes promising ideas for urban planners.

In future research, on the one hand, according to the macro applicable characteristics of e-bike, we should take consideration of different city scale and the specific investigation to understand the application of e-bike lane in diverse scenarios. On the other hand, in order to improve the accuracy of the dynamic evaluation model, we should consider the condition of limited road resource, more dimensions and the competition with various participants, especially of the conflict with motorized vehicle and pedestrian. At the same time, the model calibration data is limited to the non-motorized vehicle running characteristics in Shanghai. Nevertheless, the suitable analysis may calibrate and popularize the model according to cities with different scale.

## AUTHOR CONTRIBUTIONS

The authors confirm contribution to the paper as follows: study conception and design: L. Ling, Y. Guo; data collection: L. Ling; analysis and interpretation of results: L. Ling; draft manuscript preparation: L. Ling, X. Lai, Y. Guo, T. Tang, X. Li. All authors reviewed the results and approved the final version of the manuscript.

## ACKNOWLEDGEMENT

This study is supported by the Science and Technology Project of Zhejiang Province (2021C01011) and the Fundamental Research Funds for the Central Universities (22120210251 and 22120210252). It is also supported by the Social Science Fund of Jiangsu Province, China (No. 20GLC015) and the Natural Science Foundation of the Jiangsu Higher Education Institutions of China (No. 19KJB580003).

## REFERENCE

Cardoso, L.L., Martinez, M.C., Meléndez, A.N., Afonso, J.L., 2016. Dynamic inductive power transfer lane design for e-bikes, in: 2016 IEEE 19th International Conference on Intelligent Transportation Systems (ITSC). IEEE, pp. 2307–2312.

Chen, X., Yue, L., Han, H., 2018. Overtaking disturbance on a moped-bicycle-shared bicycle path and corresponding new bicycle path design principles. J. Transp. Eng. Part Syst. 144, 04018048.

Cherry, C.R., Yang, H., Jones, L.R., He, M., 2016. Dynamics of electric bike ownership and use in Kunming, China. Transp. Policy 45, 127–135.

Dong, S., Zhou, J., Zhao, L., Tang, K., Yang, R., 2015. Feasibility analysis of phase transition signals based on e-bike rider behavior. Adv. Mech. Eng. 7, 1687814015618905.

Electric travel, n.d. Guide ro PEV & Electric Bicycle Laws in Europe.

Guo, Y., Wang, J., Peeta, S., Anastasopoulos, P.C., 2018. Impacts of internal migration, household registration system, and family planning policy on travel mode choice in China. Travel Behav. Soc. 13, 128–143.

Haustein, S., Møller, M., 2016. E-bike safety: Individual-level factors and incident characteristics. J. Transp. Health 3, 386–394.

Huang, L., Wu, J., You, F., Lv, Z., Song, H., 2016. Cyclist social force model at unsignalized intersections with heterogeneous traffic. IEEE Trans. Ind. Inform. 13, 782–792.

Hummer, J.E., Rouphail, N.M., Toole, J.L., Patten, R.S., Schneider, R.J., Green, J.S., Hughes, R.G., Fain, S.J., 2006. Evaluation of Safety, Design, and Operation of Shared-Use Paths—Final Report.

Jamerson, F.E., Benjamin, E., 2012. Worldwide electric powered two wheel market. World Electr. Veh. J. 5, 269–275.

Ji, S., Cherry, C.R., J. Bechle, M., Wu, Y., Marshall, J.D., 2012. Electric vehicles in China: emissions and health impacts. Environ. Sci. Technol. 46, 2018–2024.

Jia, B., Li, X.-G., Jiang, R., Gao, Z.-Y., 2007. Multi-value cellular automata model for mixed bicycle flow. Eur. Phys. J. B 56, 247–252.

Jiang, R., Jia, B., Wu, Q.-S., 2004. Stochastic multi-value cellular automata models for bicycle flow. J. Phys. Math. Gen. 37, 2063.

Jin, S., Qu, X., Zhou, D., Xu, C., Ma, D., Wang, D., 2015. Estimating cycleway capacity and bicycle equivalent unit for electric bicycles. Transp. Res. Part Policy Pract. 77, 225–248.

Khan, S.I., Raksuntorn, W., 2001. Characteristics of passing and meeting maneuvers on exclusive bicycle paths. Transp. Res. Rec. 1776, 220–228.

Lai, X., Teng, J., Ling, L., 2020a. Evaluating public transportation service in a transit hub based on passengers energy cost, in: 2020 IEEE 23rd International Conference on Intelligent Transportation Systems (ITSC). IEEE, pp. 1–7.

Lai, X., Teng, J., Schonfeld, P., Ling, L., 2020b. Resilient schedule coordination for a bus transit corridor. J. Adv. Transp. 2020, 1–12.

Lee, K.H., 2004. First course on fuzzy theory and applications. Springer Science & Business Media.

Li, Y., Ni, Y., Sun, J., Ma, Z., 2020. Modeling the illegal lane-changing behavior of bicycles on road segments: Considering lane-changing categories and bicycle heterogeneity. Phys. Stat. Mech. Its Appl. 541, 123302.

Li, Z., Wang, W., Shan, X., Jin, J., Lu, J., Yang, C., 2010. Analysis of bicycle passing events for LOS evaluation on physically separated bicycle roadways in China, in: 89th Annual Meeting of the Transportation Research Board, Washington, DC.

Lin, D., Chen, X., Lin, B., Li, L., 2014. Phenomena and characteristics of moped-passing-bicycle on shared lanes, in: Proceedings of the TRB 93rd Annual Meeting Compendium of Papers. Transportation Research Board of the National Academies.

Lin, S., He, Min, Tan, Y., He, Mingwei, 2008. Comparison study on operating speeds of electric bicycles and bicycles: experience from field investigation in Kunming, China. Transp. Res. Rec. 2048, 52–59.

Ling, L., Li, F., 2017. Reliable feeder bus schedule optimization in a multi-mode transit system, in: 2017 IEEE 20th International Conference on Intelligent Transportation Systems (ITSC). IEEE, pp. 738–744.

Ling, L., Li, F., Cao, L., 2018. Analyzing the relationship between urban macroeconomic development and transport infrastructure system based on neural network, in: Green Intelligent Transportation Systems: Proceedings of the 7th International Conference on Green Intelligent Transportation System and Safety 7. Springer, pp. 763–775.

Ling, L., Murray-Tuite, P., Lee, S., Ge, Y. "Gurt", Ukkusuri, S.V., 2021. Role of Uncertainty and Social Networks on Shadow Evacuation and Non-Compliance Behavior in Hurricanes. Transp. Res. Rec. 2675, 53–64. https://doi.org/10.1177/0361198120962801

Ling, L., Qian, X., Guo, S., Ukkusuri, S.V., 2022. Spatiotemporal impacts of human activities and socio-demographics during the COVID-19 outbreak in the US. BMC Public Health 22, 1466.

Ling, L., Zhang, W., Bao, J., Ukkusuri, S.V., 2023. Influencing factors for right turn lane crash frequency based on geographically and temporally weighted regression models. J. Safety Res.

Liu, Q., Sun, J., Tian, Y., Xiong, L., 2020. Modeling and simulation of overtaking events by heterogeneous non-motorized vehicles on shared roadway segments. Simul. Model. Pract. Theory 103, 102072.

Ma, J., 2010. National Bureau of Statistics of China. China statistical yearbook. Beijing: China Statistics Press.

Manual, H.C., 2010. HCM2010. Transp. Res. Board Natl. Res. Counc. Wash. DC 1207.

Mehta, K., Mehran, B., Hellinga, B., 2015. Evaluation of the passing behavior of motorized vehicles when overtaking bicycles on urban arterial roadways. Transp. Res. Rec. 2520, 8–17.

Minh, C.C., Sano, K., Matsumoto, S., 2005. Characteristics of passing and paired riding maneuvers of motorcycle. J. East. Asia Soc. Transp. Stud. 6, 186–197.

Ning-bo, C., Zhao-wei, Q., Yong-heng, C., Li-ying, Z., Xian-min, S., Qiao-wen, B., 2017. Destination and route choice models for bidirectional pedestrian flow based on the social force model. IET Intell. Transp. Syst. 11, 537–545.

Qu, Z., Cao, N., Chen, Y., Zhao, L., Bai, Q., Luo, R., 2017. Modeling electric bike–car mixed flow via social force model. Adv. Mech. Eng. 9, 1687814017719641.

Salmeron-Manzano, E., Manzano-Agugliaro, F., 2018. The electric bicycle: Worldwide research trends. Energies 11, 1894.

Shan, X., Li, Z., Chen, X., Ye, J., 2015. A modified cellular automaton approach for mixed bicycle traffic flow modeling. Discrete Dyn. Nat. Soc. 2015.

Tang, T., Guo, Y., Zhou, X., Labi, S., Zhu, S., 2021. Understanding electric bike riders' intention to violate traffic rules and accident proneness in China. Travel Behav. Soc. 23, 25–38.

Ukkusuri, S., Ling, L., Le, T.V., Zhang, W., 2020. Performance of Right-Turn Lane Designs at Intersections.

Vinay, S., Aithal, S., Sudhakara, G., 2013. A quantitative approach using goal-oriented requirements engineering methodology and analytic hierarchy process in selecting the best alternative, in: Proceedings of International Conference on Advances in Computing. Springer, pp. 441–454.

Virkler, M.R., Balasubramanian, R., 1998. Flow characteristics on shared hiking/biking/jogging trails. Transp. Res. Rec. 1636, 43–46.

Wang, C., Xu, C., Xia, J., Qian, Z., 2018. The effects of safety knowledge and psychological factors on self-reported risky driving behaviors including group violations for e-bike riders in China. Transp. Res. Part F Traffic Psychol. Behav. 56, 344–353.

Wang, T., Chen, J., Wang, C., Ye, X., 2018. Understand e-bicyclist safety in China: Crash severity modeling using a generalized ordered logit model. Adv. Mech. Eng. 10, 1687814018781625.

World Health Organization, n.d. Protecting Chinese e-bike users from road injuries and deaths.

Xu, L., Liu, M., Song, X., Jin, S., 2018. Analytical model of passing events for one-way heterogeneous bicycle traffic flows. Transp. Res. Rec. 2672, 125–135.

Zadeh, L.A., Klir, G.J., Yuan, B., 1996. Fuzzy sets, fuzzy logic, and fuzzy systems: selected papers. World Scientific.

Zhao, D., Wang, W., Li, C., Li, Z., Fu, P., Hu, X., 2013. Modeling of passing events in mixed bicycle traffic with cellular automata. Transp. Res. Rec. 2387, 26–34.

Zheng, C., Shen, J., Zhang, Y., Huang, W., Zhu, X., Wu, X., Chen, L., Gao, X., Cen, K., 2017. Quantitative assessment of industrial VOC emissions in China: Historical trend, spatial distribution, uncertainties, and projection. Atmos. Environ. 150, 116–125.

Zhou, D., Jin, S., Ma, D., Wang, D., 2015. Modeling mixed bicycle traffic flow: a comparative study on the cellular automata approach. Discrete Dyn. Nat. Soc. 2015.